\documentclass{svjour3}                     % onecolumn (standard format)
\smartqed  % flush right qed marks, e.g. at end of proof
\usepackage{amsmath,amssymb}
\usepackage[dvips]{graphicx,color,epsfig}
\usepackage{subfigure}
\usepackage{float}

\newcommand{\p}{\partial}

\renewcommand{\Re}{\mathop{\rm Re}}
\newcommand{\ka}{\hbox{\ae}}
\DeclareMathOperator{\sech}{sech}
 \journalname{Optical and Quantum Electronics}
\begin{document}

\title{Second harmonic generation and pulse shaping in positively and negatively spatially dispersive nanowaveguides: comparative analysis%\thanks{Grants or other notes
}

\titlerunning{SHG in negatively dispersive nanowaveguides}        % if too long for running head

\author{Alexander K. Popov         \and
        Sergey A. Myslivets %etc.
}


\institute{Alexander K. Popov \at
              Birck Nanotechnology Center\\
Purdue University\\
West Lafayette, Indiana 47907--2057\\
Email: popov@purdue.edu,\\ https://nanohub.org/groups/nlo/popov           %  \\
           \and
           Sergey A. Myslivets \at
             Institute of Physics, Siberian Branch of the Russian Academy of Sciences\\ and Siberian Federal University \\
660036 Krasnoyarsk, Russia\\
Email: sam@iph.krasn.ru
}

\date{Received: date / Accepted: date}

\maketitle

\begin{abstract}
Comparative analysis of second harmonic generation in ordinary and backward-wave settings is presented. Extraordinary properties of frequency doubling nonlinear optical reflectivity and pulse shaping through phase matching of ordinary and backward electromagnetic waves in the nanowaveguides with mixed negative/positive spatial dispersion is demonstrated with numerical simulations.
\end{abstract}

\section{Introduction}
\label{intro}

Metamaterials (MM) are artificially designed and engineered materials, which can have properties unattainable in nature. Usually, MM rely on advances in nanotechnology to build tiny metallic nanostructures smaller than the wavelength of light. These nanostructures modify the electromagnetic (EM) properties of MM, sometimes creating seemingly impossible optical effects. Negative-index MM (NIMs) are the most intriguing EM materials that support backward EM waves (BEMW). Phase velocity and energy flux (group velocity) become \emph{contra-directed} in NIMs. The appearance of BEMW is commonly associated with simultaneously negative electric permittivity ($\epsilon<0$) and magnetic permeability ($\mu<0$) at the corresponding frequency and, consequently, with negative refractive index $n= - \sqrt{\mu\epsilon}$. Counter-intuitive backwardness of EMW is in strict contrast with the electrodynamics of ordinary, positive-index materials. Extraordinary properties of nonlinear-optical frequency converting propagation processes has been predicted provided that one of the coupled electromagnetic waves falls in the negative-index frequency domain and propagates against the others \cite{1,2,3,4}.
Among them are second harmonic generation  (SHG) \cite{1,2,3,4}, optical parametric amplification and frequency-shifted nonlinear reflectivity \cite{1,2,3,4,5}.

Current mainstream in fabricating bulk NIM slabs relies on engineering of LC nanocircuits - plasmonic mesoatoms with phase shifted, negative electromagnetic response. Extraordinary coherent, nonlinear optical, frequency converting propagation processes predicted in NIMs have been experimentally realized so far only in the microwave \cite{6}. A different paradigm which employs spatial dispersion \cite{AgGa,Agr} was proposed to realize outlined processes. Basic underlaying idea is nanoengineering negative spatial dispersion which would enable negative group velocity $v_{gr}=\partial\omega/\partial k<0$. Then energy flux, which is directed along the group velocity, appears directed \emph{against} the wave vector. Such a property makes an analog of the effective negative refractive index.  Various particular realizations of negative dispersion were proposed \cite{Tr,NT,APA2012,AST2013,SSP2014}. Basically, many hyperbolic metamaterials  may support backward-wave  electromagnetic modes, see, e.g., \cite{Nar}, as well as specially designed waveguides \cite{Mok,Chr}. This opens new avenues for creation of backward-wave photonic devices with unparalleled functional properties. As regards coherent nonlinear-optical propagation processes, critically important is to provide for coexistence of ordinary and backward coupled waves which frequencies and wave vectors satisfy to energy and momentum conservation law (phase matching). Such possibilities were described in ref.~\cite{APA2012,AST2013,SSP2014}. Corresponding MM slabs can be viewed as the plasmonic nanowaveguides which support both ordinary EMWs with co-directed phase velocity and energy flux ($\partial\omega/\partial k>0$) and extraordinary, BW  modes with contra-directed phase and group velocities ($\partial\omega/\partial k<0$). Some of them exhibit spatial dispersion which changes from  positive to negative in different wavelength intervals whereas the others predominantly maintain negative  dispersion. Frequencies and gaps between the modes can be tailored by changing shapes, sizes and spacing of the MM nano-building blocks  so that phase velocities for the modes matching the photon energy conservation law would become equal as their group velocities remain contra-directed. Basically, different materials and geometries can be utilized to control spatial dispersion and losses in such MM with variable spatial dispersion.

Second harmonic generation (SHG) is the basic fundamental coherent (i.e., phase-dependent) nonlinear optical propagation process which finds many practically important application. Overall conversion efficiency is controlled by the intensity dependent local conversion rate and by the corresponding depletion of the fundamental beam.  As shown in \cite{1,2,3,4,5}, such process exhibit unusual properties in the BW settings which are significantly different from both BW optical parametrical amplification and SHG in ordinary materials.   This paper is to analyze such properties of SHG and fundamental differences between  SHG in ordinary and backward-wave settings with focus on the pulse regime which is of  particular interest.
\section{Basic Equations}\label{pul}
As noted, the properties of SHG will be investigated  for the cases where one of the coupled waves is ordinary and the other one is backward wave, in continuous-wave and pulse regimes, and compared with their ordinary counterparts.  To achieve phase matching, phase velocities of the fundamental  the SH waves must be co-directed. This means that in BW setting and pulse regime, the pulse of SH  will propagate \emph{against} the pulse of the fundamental radiation. Such extraordinary NLO coupling scheme dictates {exotic} properties of the \emph{ frequency-doubling  ultracompact NLO mirror} under consideration, which are described below.
One of our goals is to investigate  the requirements for the magnitude of the metamaterial nonlinearity as well as for intensity of the control field(s) that would enable efficient functional properties of the indicated SHG mirror. Corresponding basic equations are as follows.

Electric and magnetic components of the waves and corresponding nonlinear polarizations are defined as
\begin{eqnarray}
\{{\mathcal{E}}, {\mathcal{H}}\}_j=\Re{\{{E},{H}\}_j\exp\{i(k_jz-\omega_jt)\}},\label{eh}&&\\
%\mathcal{E}_j=\Re{E_j\exp\{i(k_jz-\omega_jt)\}},\label{m}&&\\
\{\mathcal{P}, \mathcal{M}\}^{NL}_j=\Re{\{P, M\}^{NL}_j\exp\{i(\widetilde{k}_jz-\omega_jt)\}},\label{pme}&&\\
%\mathcal{P}^{NL}_j=\Re{P^{NL}_j\exp\{i(\widetilde{k}_jz-\omega_jt)\}},\label{pe}&&\\
\{P, M\}^{NL}_1=\chi^{(2)}_{e,m,1}\{E, H\}_2\{E, H\}_1^*,\label{pm1}&&\\
 \{P, M\}^{NL}_2=\chi^{(2)}_{e,m,2}\{E, H\}_1^2,\, 2\chi^{(2)}_{e,m,2}=\chi^{(2)}_{e,m,1}. \label{pm2}&&%\\
%M^{NL}_1=\chi^{(2)}_{m,1}H_2H_1^*,\, M^{NL}_2=\chi^{(2)}_{m,2}H_1^2, \, 2\chi^{(2)}_{m,2}=\chi^{(2)}_{m,1}. \label{ppm}&&
\end{eqnarray}
Equations for amplitudes $E_j$ can be written as
\begin{eqnarray}
s_2\frac{\p E_2}{\p z}+ \frac1{v_2}\frac{\p E_2}{\p t}= - i\frac{k_2\omega_2^2}{\epsilon_2c^2}4\pi\chi^{(2)}_{e,2}E_1^2\exp{(-i\Delta kz)}-\frac{\alpha_2}2E_2,\label{eq1a}&&\\
  s_1\frac{\p E_1}{\p z}+ \frac1{v_1}\frac{\p E_1}{\p t}= - i\frac{k_1\omega_1^2}{\epsilon_1c^2}8\pi\chi^{(2)*}_{e,1}E_1^*E_2\exp{(i\Delta kz)}-\frac{\alpha_1}2E_1.\label{eq1b}&&
\end{eqnarray}
Here,  $v_i>0$ and $\alpha_{1,2}$  are  group velocities and absorption indices at the corresponding frequencies, $\chi^{(2)}_{\rm eff}=\chi^{(2)}_{e,2}$ is  effective nonlinear susceptibility, $\Delta k=k_{2}-2k_{1}$.   Parameter $s_j=1$ for ordinary, and $s_j=-1$ for backward wave.
With account for $k^2=n^2(\omega/c)^2$, $n_1=s_1\sqrt{\epsilon_1\mu_1}$, $n_2=s_2\sqrt{\epsilon_2\mu_2}$, we introduce
amplitudes $e_{j}=\sqrt{|\epsilon_j|/k_j}E_j$,  $a_j=e_i/e_{10}$, coupling parameters $\ka=\sqrt{k_1k_2/|\epsilon_1\epsilon_2|} 4\pi\chi^{(2)}_{\rm eff}$ and $g=\ka E_{10}$, loss  and phase mismatch parameters $\widetilde \alpha_{1,2}=a_{1,2}L$ and  $\Delta \widetilde{k}=\Delta {k}l$, slub thickness $d=L/l$, position $\xi=z/l$ and time instant $\tau=t/\Delta\tau$. It is assumed that   $E_{j0}=E_j(z=0)$, $l=v_1\Delta\tau$ is the pump pulse length, $\Delta\tau$ is duration of the input fundamental pulse. Quantities  $|a_j|^2$ are proportional to the time dependent photon fluxes. Then Eqs. \eqref{eq1a} and \eqref{eq1b} are written as
\begin{eqnarray}
s_2\frac{\p a_2}{\p \xi}+ \frac{v_1}{v_2}\frac{\p a_2}{\p \tau}=
  -igla_1^2\exp{(-i\Delta \widetilde{k}\xi)}-\frac{\widetilde{\alpha}_2}{2d}a_2, \label{eq2a}&&\\
    s_1\frac{\p a_1}{\p \xi}+\frac{\p a_1}{\p \tau}=
     -i2g^*la_1^*a_2\exp{(i\Delta \widetilde{k}\xi)}-\frac{\widetilde{\alpha}_1}{2d}a_1. \label{eq2b}&&
\end{eqnarray}
In the case of magnetic nonlinearity, $\chi^{(2)}_{\rm eff}=\chi^{(2)}_{m,2}$, equations for the magnetic components of the fields $m_{j}=\sqrt{|\mu_j|/k_j}H_j$,  $a_j=m_i/m_{10}$, take the form of \eqref{eq2a} and \eqref{eq2b} where coupling parameters are $\ka=\sqrt{k_1k_2/|\mu_1\mu_2|} 4\pi\chi^{(2)}_{\rm eff}$ and $g=\ka  H_{10}$. In both cases, $z_0=g^{-1}$ is characteristic medium length required for significant NLO energy conversion. Only most favorable case of the exact phase matching $\Delta {k}=0$ and loss-free medium ($\alpha_1=\alpha_2=0$) will be considered below.

\section{Phase matched SHG of continuous wave in loss-free medium}
For the case of $l\rightarrow\infty$, ${\p a_j}/{\p \tau}\rightarrow 0$ and $\widetilde \alpha_{1,2}=\Delta \widetilde{k}=0$, Eqs.~\eqref{eq2a} and \eqref{eq2b} reduce to
\begin{equation}
 s_2{d a_2}/{d z} = -iga_1^2, \quad
    s_1{d a_1}/{d z}= -i2g^*a_1^*a_2. \label{eqcw}
\end{equation}
Manley-Rove equation (photon conservation law) derived from Eqs.\eqref{eqcw} is
\begin{equation}
 s_2{d |a_2|^2}/{d z} + (1/2)s_1{d |a_1|^2}/{d z}=0. \label{mr}
\end{equation}
{\bf \subsection{SHG in ordinary NLO medium}}
For the case of ordinary waves $s_1=s_2=1$ and $a_{20}=0$ with account for $a_{10}=1$, one finds from Eq.~\eqref{mr}  that
\begin{equation}
2|a_2|^2+|a_1|^2=1. \label{pimmr}
\end{equation}
Solution to  Eqs. \eqref{eqcw} is found as \cite{2,3}
\begin{equation}
2|a_2|^2=\tanh^2{(\sqrt{2}gz)},\quad
|a_1|^2=\sech^2{(\sqrt{2}gz)}.  \label{shgpim}
\end{equation}
\begin{figure}[!h]
          \centering
         \subfigure[]{\includegraphics[width=.24\textwidth]{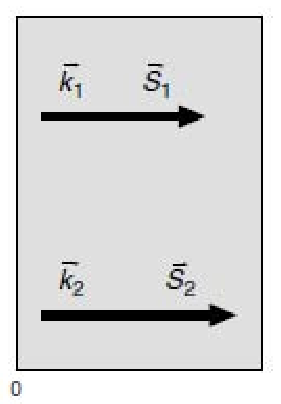}}
         \subfigure[]{\includegraphics[width=.46\textwidth]{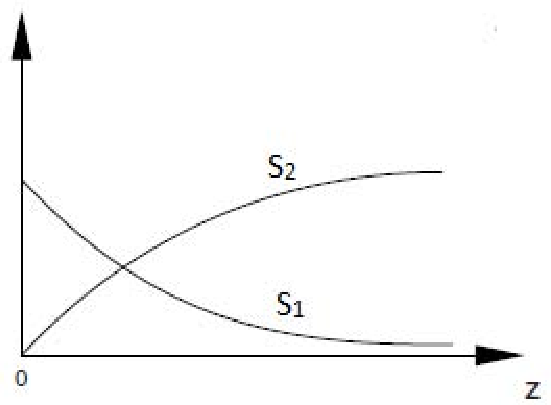}}
          \caption{Coupling geometry and energy fluxes in ordinary, positive index, materials.}\label{PIM}
\end{figure}
Figures \ref{PIM} (a) and (b) schematically show coupling geometry and  energy fluxes  $S_{2,1}\propto |a_{2,1}|^2 $ across the ordinary, positive index material slab. According to Eqs.~\eqref{pimmr} and \eqref{shgpim}, \emph{sum} of fluxes of photons $\hbar\omega_2$  and of the pairs of photons $\hbar\omega_1$ is conserved across the medium.
{\bf \subsection{SHG in the nanowaveguide with mixed, negative   at $\omega$ and positive at $2\omega$ spatial dispersion}}
Figures~\ref{NIM} (a) and (b) display coupling geometry  and energy fluxes in the NLO slab in this case.
\begin{figure}[!h]
          \centering
         \subfigure[]{\includegraphics[width=.21\textwidth]{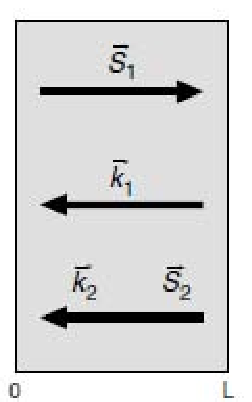}}
         \subfigure[]{\includegraphics[width=.46\textwidth]{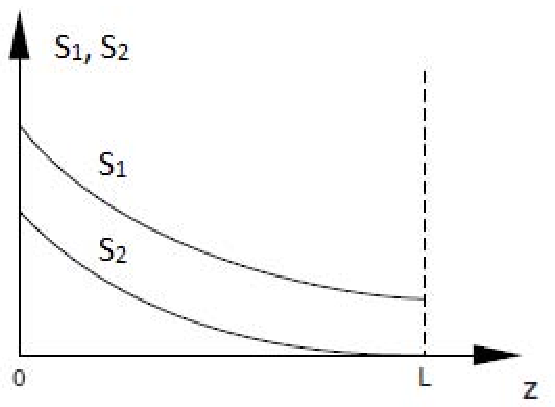}}
          \caption{Coupling geometry and energy fluxes in negative-index metamaterials.}\label{NIM}
\end{figure}
In the case of backward fundamental  and ordinary SH waves, $s_1=-1, s_2=1$,   equations \eqref{eqcw} and \eqref{mr} dictates fundamentally different behavior. Equation \eqref{mr} predicts
\begin{equation}
|a_1|^2-2|a_2|^2=|B|^2, \label{nimmr}
\end{equation}
where $|B|^2$ is a constant which, however,  depends on the slab thickness and on strength of the input fundamental field.  Equations~\eqref{eqcw} reduce to
\begin{equation}
{d a_2}/{d z} = -iga_1^2, \quad
    {d a_1}/{d z}= i2g^*a_1^*a_2. \label{eqnim}
\end{equation}
Besides the fact that the in this case the equations have different signs in the  right sides, the boundary conditions for fundamental and SH waves must be applied to \emph{opposite} edges of the slabs of thickness $L$: $a_{10}=1$, $a_{2L}=0$. Indicated differences give rise to fundamental changes in the solution to the equations for the amplitudes and, consequently, to the properties of the SHG as a whole. Solution to Eqs.~\eqref{eqnim} can be presented as \cite{2,3}
\begin{eqnarray}
\sqrt{2}a_2 = B\tan{[B\sqrt{2}g(L-z)]},\label{shgnim2}&&\\
a_1=B\sec{[B\sqrt{2}g(L-z)]}. \label{shgnim1}&&
\end{eqnarray}
 As follows from \eqref{shgnim1}, quantity $B$ presents transmitted fundamental wave, $a_{1L}=B$. It is  found from Eq.~\eqref{shgnim1} at $z=0$ as:
\begin{equation}
B= \cos{(B\sqrt{2}gL)}.\label{B}
\end{equation}

According to Eqs.~\eqref{nimmr} and \eqref{shgnim2} --  \eqref{shgnim1}, \emph{difference} between the numbers of pairs of photons $\hbar\omega_1$ and the number of photons $\hbar\omega_2$ is conserved so that the number of photons $\hbar\omega_2$ at $z=L$ is always equal to zero. The later requirement controls whole process of BWSHG. Such a dependence is in strict contrast with SHG in ordinary NLO materials, where the SH flux may exceed the fundamental one in some area inside the slab. In the case of BWSHG, its energy flux is directed against the fundamental one towards the areas with higher intensity of the fundamental wave whereas remains lower everywhere inside the slab. As seen from Eq.~\eqref{B}, the gap between the two curves, $|B|^2$, decreases and, hence, the conversion efficiency grows with $gL=L/z_0 \rightarrow \infty$. Like in the preceding case, $z_0$ represent a characteristic medium length required for significant energy conversion to the SH. It shortens with growth of strength of the pump field.

\begin{figure}[!ht]
          \centering
        \subfigure[]{\includegraphics[width=.45\textwidth]{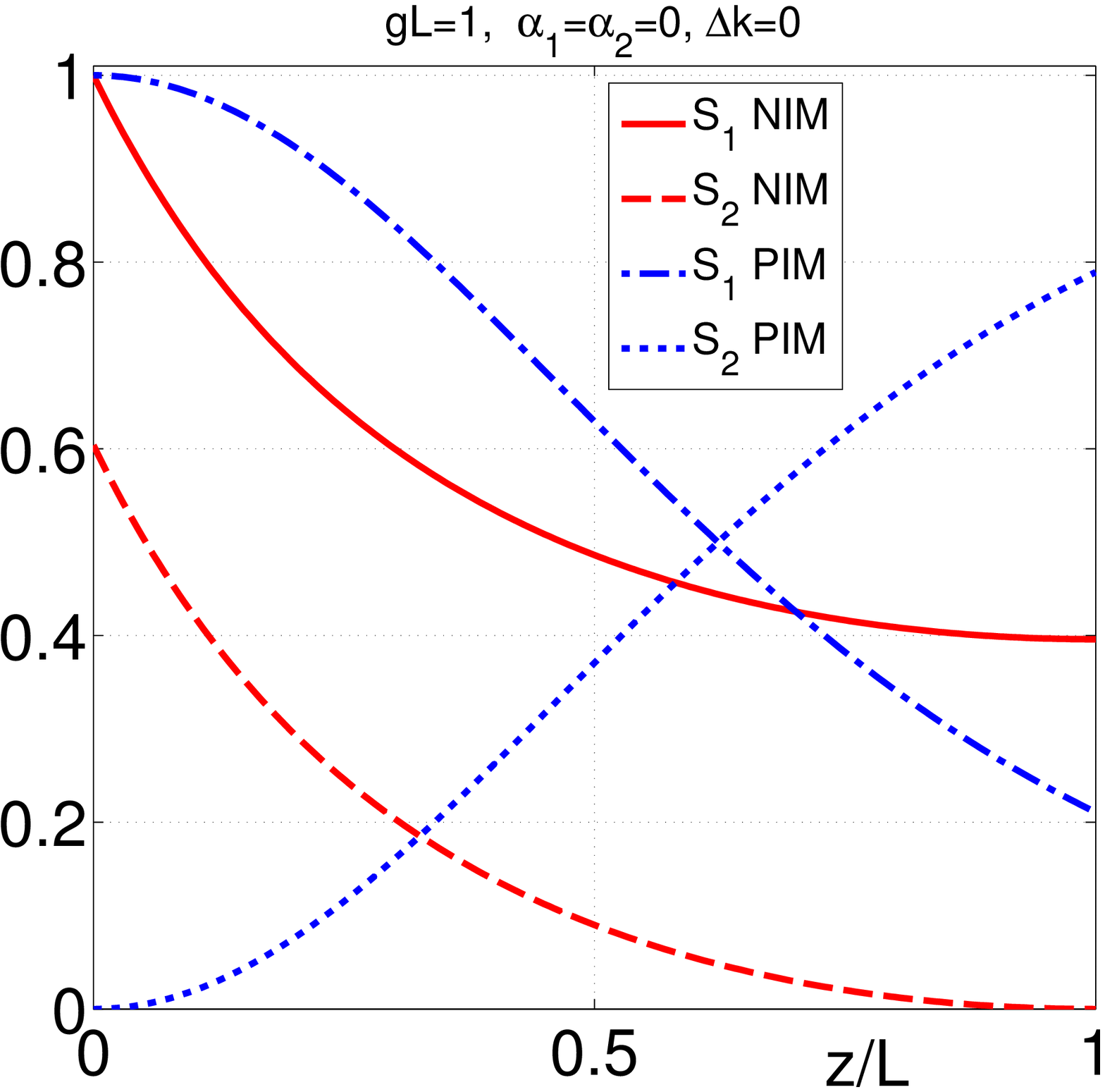}}
     \subfigure[]{\includegraphics[width=.45\textwidth]{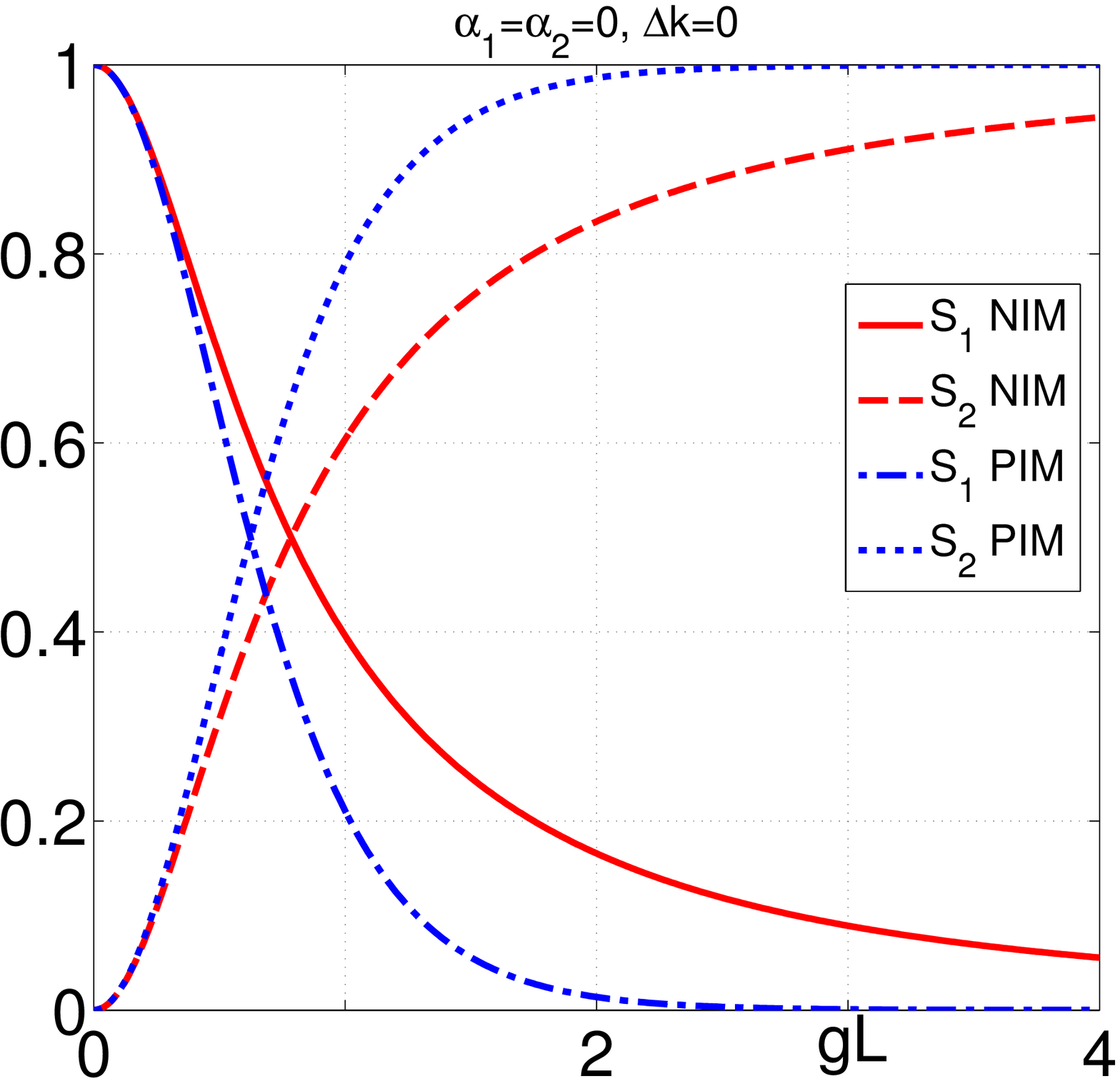}}
          \caption{Differences between  SHG in ordinary and BW settings. (a). {Energy fluxes across the slab for fundamental (the descending dash-dotted blue line) and SH (the ascending dotted blue line), both are ordinary waves in a positive-index (PIM) material; and for \emph{backward-wave} fundamental (the descending solid red line) and ordinary-wave SH (the \emph{descending} dashed red line)  in a negative-index (NIM) slab. $gL=1$.} (b).{Output transmitted fundamental (the descending blue dash-dotted line) and  SH (the ascending blue dotted line) at $z=L$, both are ordinary waves in a PIM; and transmitted backward-wave fundamental (the descending solid red line) flux at $z=L$  and   ordinary  SH flux at $z=0$ (the ascending dashed  red line) in a NIM.}}\label{NIMPim}
\end{figure}

{\bf \subsection{Comparison of  SHG for ordinary and backward fundamental waves}}

Figure \ref{NIMPim}(a) illustrate unparalleled properties of SH generation with backward waves as compared with its ordinary counterpart at  similar other parameters for the particular example of $gL=1$. As noted, the remarkable property is the fact that SH propagates against the fundamental wave and, therefore,  metaslabs operates as a \emph{frequency doubling metamirror} with reflectivity controlled by the fundamental wave. Further materials  which support only ordinary waves and MM that support both ordinary and backward waves will be referred to as PIM and NIM.
Figure \ref{NIMPim}(b) compares output intensity of the SH and transmitted fundamental wave. Major important conclusion are as follows. It appears that the efficiency of SHG in ordinary settings exceeds that in the NIM metaslab at equal other parameters. At that, propagation properties of SH  appear fundamentally different which holds promise of extraordinary applications. Intensity of SH is less than  that of the fundamental wave   across the NIM slab, whereas it can exceed that in the vicinity of exit from  PIM slab [Fig.~\ref{NIMPim}(a)] Quantum conversion grows sharper with increase of intensity of fundamental beam and higher conversion at lower intensities occurs in PIM as compared with NIM.
\\%
\section{Comparison of ordinary and BW SHG in short-pulse regime}
In this section, comparative analysis of SHG  in the ordinary and BW settings will be done for the case where intensity of input fundamental wave varies in time and, therefore, SHG is described  by the system of partial differential equations (\ref{eq2a}) and  (\ref{eq2b}).
The input pulse shape is chosen close to a rectangular form
\begin{equation}
F(\tau)=0.5\left(\tanh\frac{\tau_0+1-\tau}{\delta\tau}-\tanh\frac{\tau_0-\tau}{\delta\tau}\right),
\end{equation}
where $\delta\tau$ is the duration of the pulse front and tail, and $\tau_0$ is the shift of the front relative to $t=0$.
Parameters $\delta\tau=0.01$ and $\tau_0=0.1$ have been selected for numerical simulations. Absorption is neglected ($\alpha_1=\alpha_2=0$). Phase velocities are equal $\Delta k=0$. Group velocities are equal (${\rm v_{12}}$) and contra-directed.

 %
 %%%%%%%%%%%%%%%%%%%%%%%%%%%%%%%%%
 \begin{figure}[hbt]
          \centering
        \subfigure[]{\includegraphics[width=.45\textwidth]{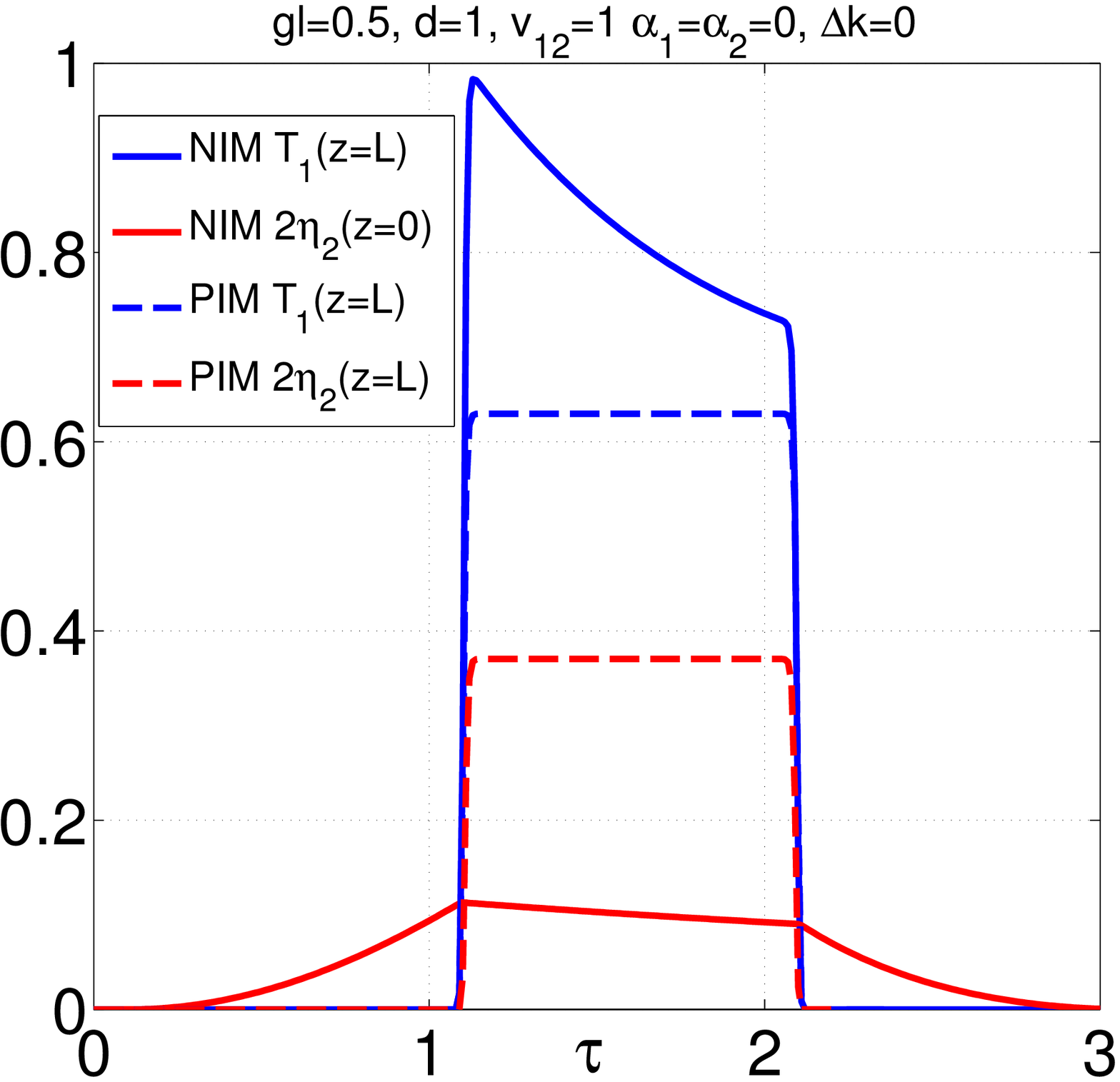}}
        \subfigure[]{\includegraphics[width=.45\textwidth]{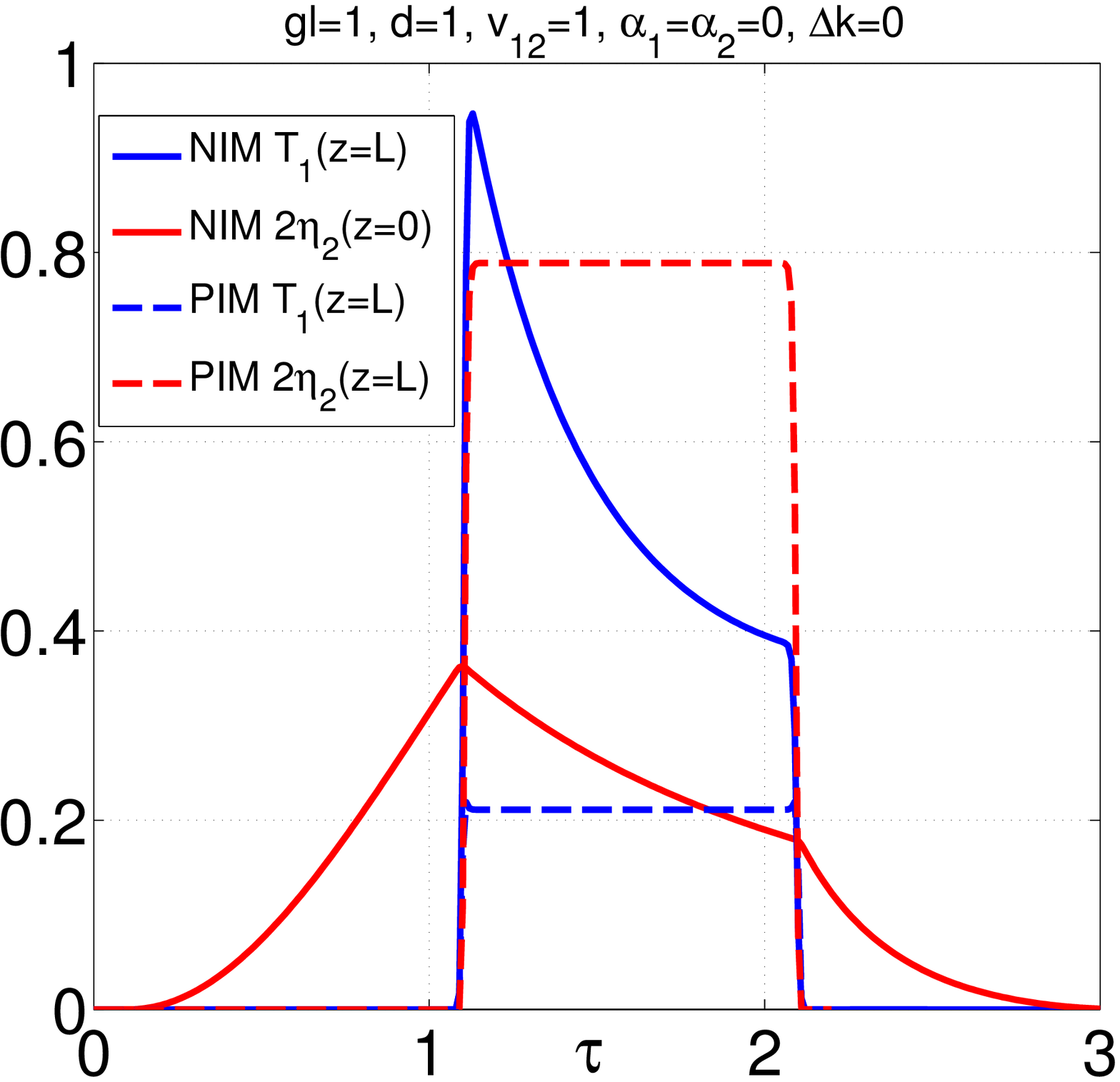}}\\
          \subfigure[]{\includegraphics[width=.45\textwidth]{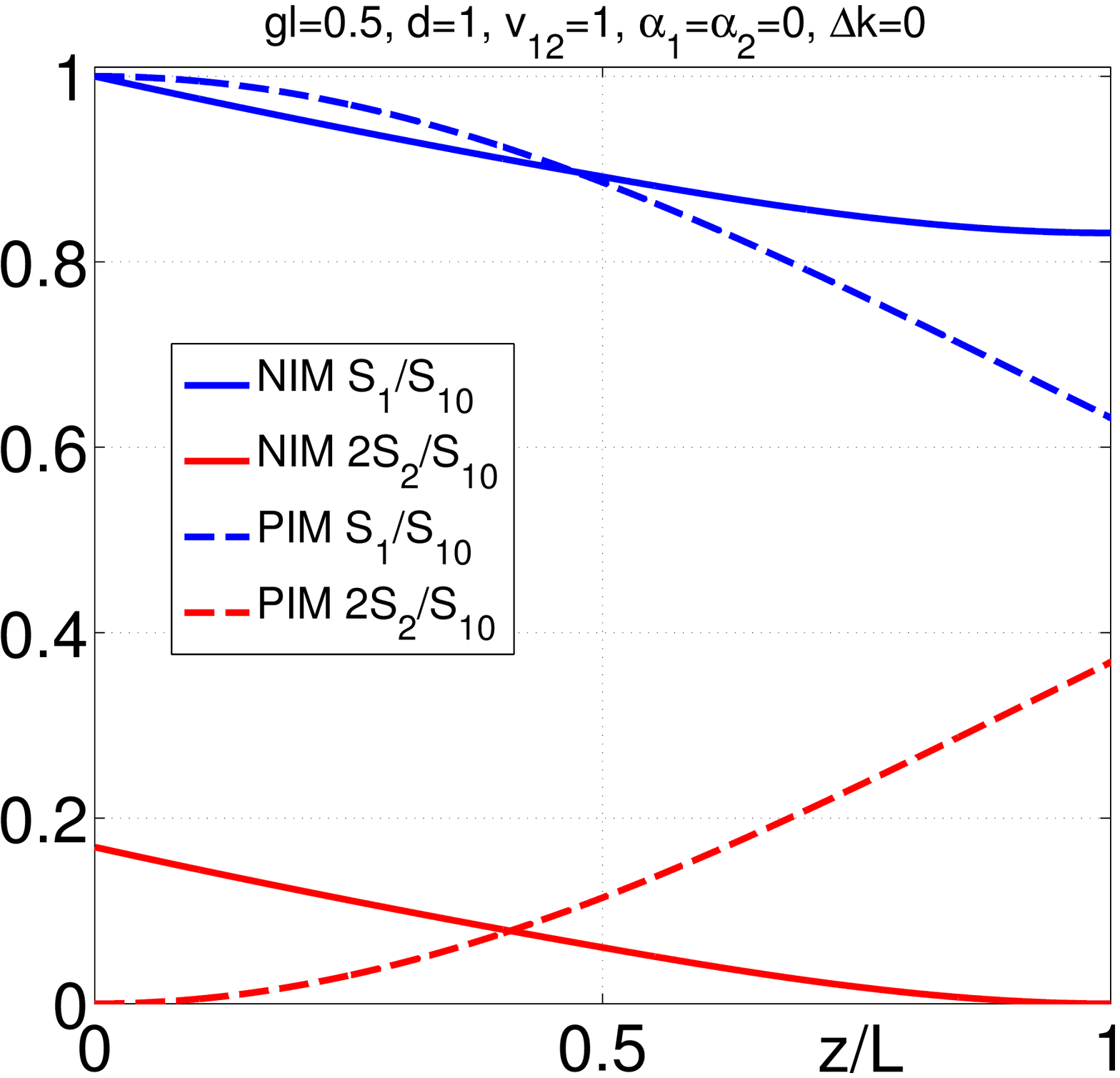}}
           \subfigure[]{\includegraphics[width=.45\textwidth]{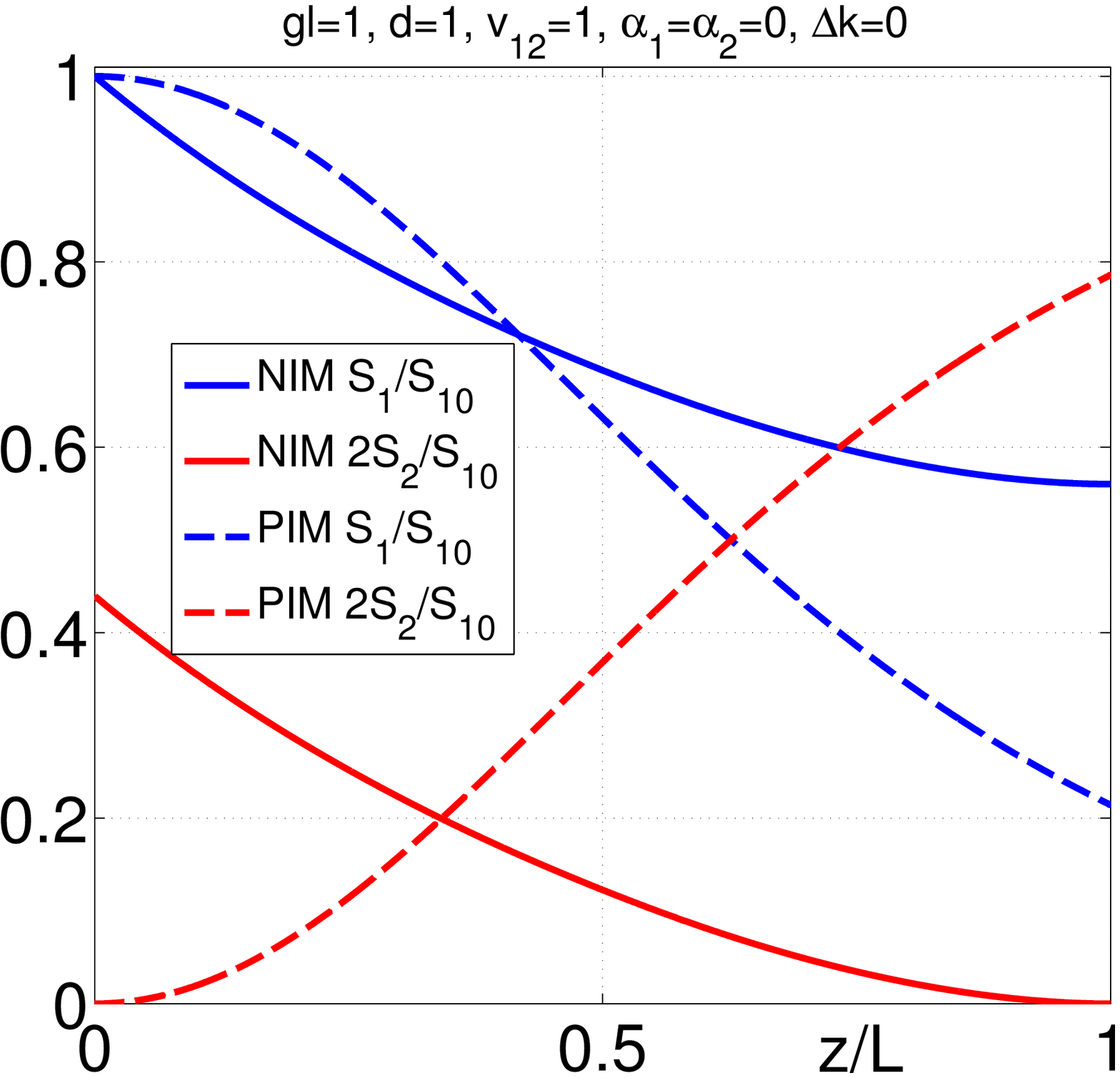}}
                   \caption{Comparison of pulse shapes and energy conversion at SHG in ordinary (PIM) and backward-wave (NIM) settings in a loss-free MM. The length of input pulse at the fundamental frequency is equal to the metaslab thickness. (a,b): Input rectangular $T_1$  pulse shapes for the fundamental radiation; $\eta_2$ -- for SH.  (c,d): Change of the pulse energy at the corresponding frequencies across the slab. Here, $d=L/l$, $S_1(z)$ and $S_{10}=S_1(z=0)$ are fundamental pulse energy, $2S_2/S_{10}$ is energy (photon) conversion efficiency per pulse.}\label{fig4}
\end{figure}

 \begin{figure}[!h]
          \centering
                    \subfigure[]{\includegraphics[width=.45\textwidth]{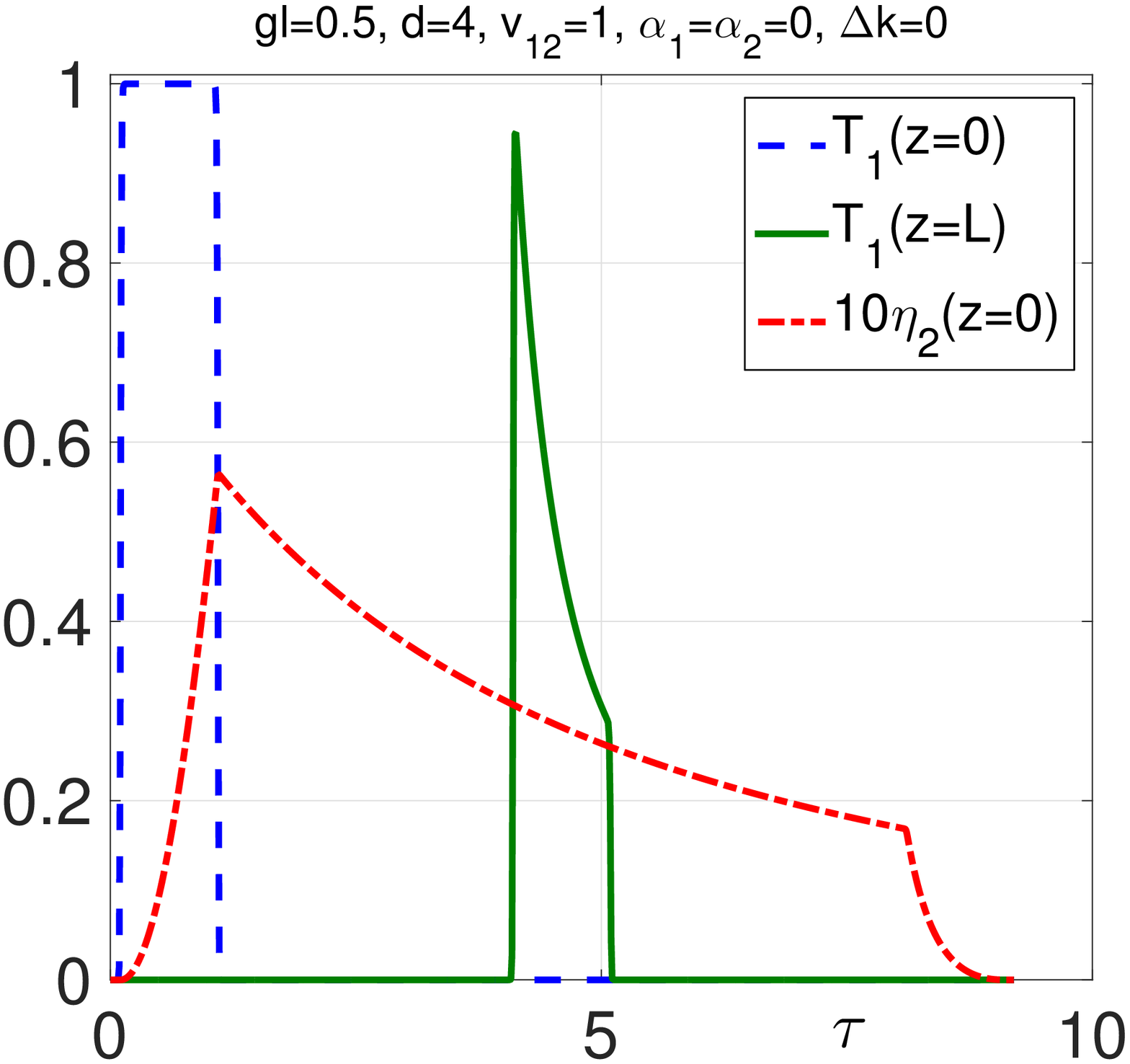}}
                    \subfigure[]{\includegraphics[width=.45\textwidth]{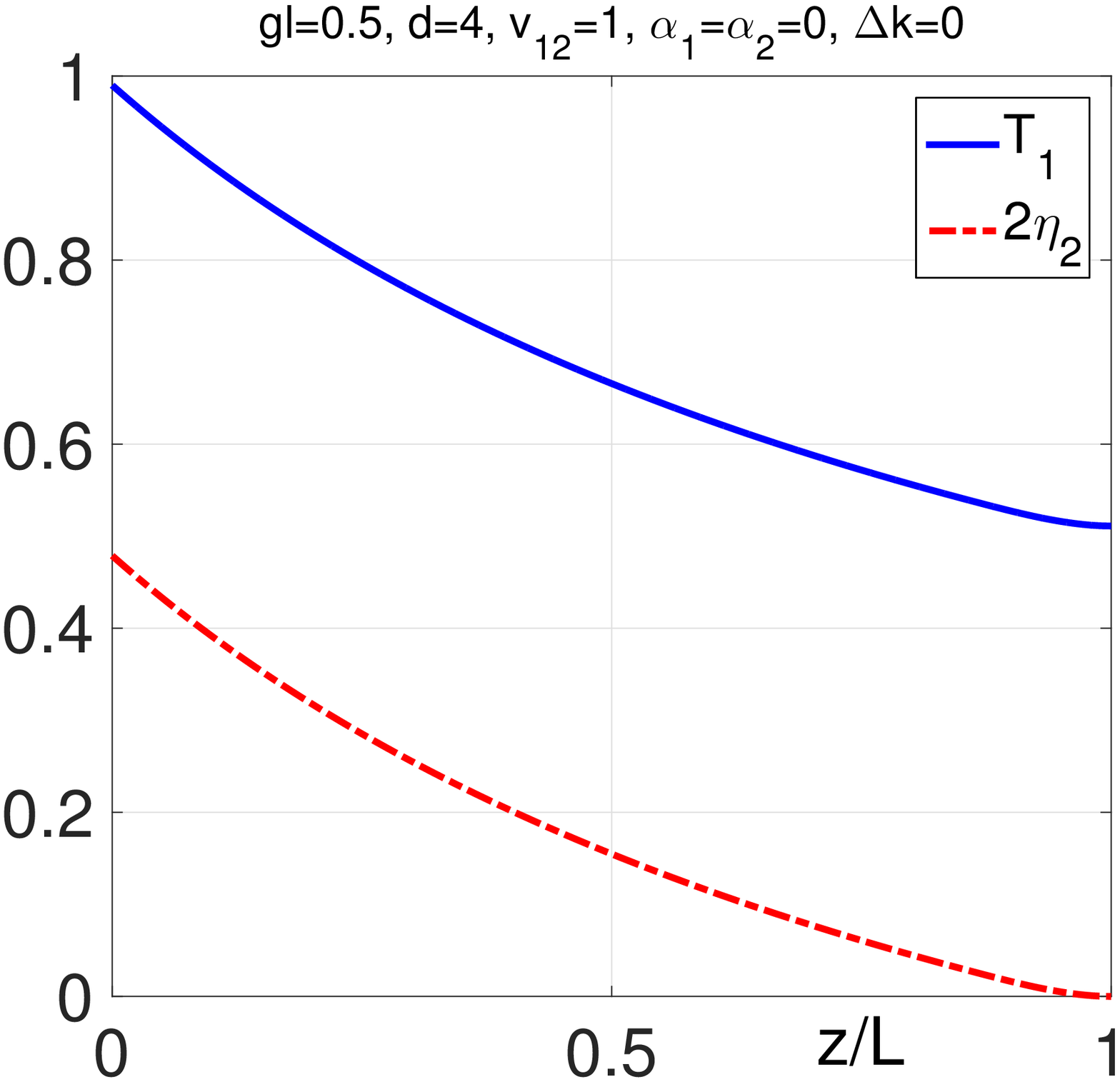}}
          \caption{Effect of pulse width. Here input pulse duration is decreased in four times as compared with Fig.~\ref{fig4} (a) at the same peak intensity. $T_1$  is pulse shape for the fundamental radiation, $\eta_2$ -- for SH,  $S_1(z)$ and $S_{10}=S_1(z=0)$ are fundamental pulse energy, $2S_2/S_{10}$ is energy (photon) conversion efficiency per pulse,  $d=L/l$,}\label{fig5}
\end{figure}

 Unusual properties of SHG in NIMs in the pulsed regime stem from the fact that it occurs only inside the traveling pulse of fundamental radiation. Generation begins on its leading edge, grows towards its trailing edge, and then exits the fundamental pulse with no further changes. Since the fundamental pulse propagates across the slab, the duration of the SH pulse may occur significantly \emph{longer} than that of the fundamental one. Depletion of the fundamental radiation along the pulse and the overall conversion efficiency depend not only on maximum intensity of the input pulse, on matching of phase and group velocities of the fundamental and second harmonic, but on the ratio of the fundamental pulse  length and slab thickness.
 Such properties are in strict contrast with that of SHG in PIM as illustrated in Figs.~\ref{fig4} and \ref{fig5}.
  Shape of the input fundamental pulse is given by the function $T_1=|a_1(\tau, z=0)|^2/|a_{10}|^2$  when its leading front enters the medium. The results of numerical simulations  for the output fundamental pulse, when its tail reaches the slab boundary, is  given by $T_1=|a_1(\tau, z=L)|^2/|a_{10}|^2$. Shape of the output pulse of SH,  when its tail passes the slab's edge at $z=0$, are  given by the function  $\eta_2=|a_2(\tau, z=0)|^2/|a_{10}|^2$. Pulse energy are represented by the time integrated pulse areas $S_j$ which varies across the slab.  As seen from Figs.~\ref{fig4} (a) and (b), saturation is homogeneous across the pulse in PIM and shape of both fundamental and SH output pulses remain rectangular. On the contrary,  shapes of the output fundamental and SH pulses in PIM are \emph{different} and \emph{change} with change of intensity of input fundamental pulse. Moreover, it appears that shapes of the output pulses varies with change of the input pulse length at unchanged  other parameters  as seen from Figs.~\ref{fig5} (a). Basic properties of pulse energy conversion across the NIM and PIM  slabs  qualitatively  resemble those in continuous-wave regime [Figs.~\ref{fig4} (c) and (d)]. \emph{Unparallel} property in NIM is growth of pulse energy conversion with shortening of pulse length at it constant instant intensity [cf. Figs.~\ref{fig4} (c) and \ref{fig5} (b)].
Figures \ref{fig5}(a) and (b) correspond to the fundamental pulse four time shorter than the slab thickness. They show an increase of the conversion efficiency  with increase of intensity of the input pulse. It is followed by the shortening of the SH pulse.

Figures ~\ref{fig4} (c), (d) and \ref{fig5} (b)  satisfy to the conservation law in a loss-free metaslab: the number of annihilated pair of photons of fundamental radiation $(S_{10}-S_{1L})/2$ is equal to the number of output SH photons $S_{20}$.
 Figures~\ref{fig4}~(a),~(b) and \ref{fig5}(a) prove that shapes and width of fundamental and generated SH pulses as well as the energy conversion efficiency to the reflected pulses at doubled frequency  can be controlled by changing intensity and ratio of the the input pulse length to the metamaterial thickness (parameter $d$).

\section{Conclusion}
Comparative analysis of second harmonic generation in ordinary and backward-wave settings in continuous and pulse regimes is presented. Backward-wave regime is attributed to the nanomaterials which possess spatial dispersion of opposite signs  at fundamental at second harmonic frequencies. This makes possible equal phase velocities of the coupled waves whereas their energy fluxes appear contra-directed. Such deliberately engineered nanolayers can be viewed as nanowaveguides. It is shown that properties of second harmonic generation in ordinary and backward-wave settings are fundamentally different. In the latter case, metaslab serves as microscopic frequency doubling nonlinear-optical mirror which properties can be all-optically controlled. In pulse regime, energy conversion property are qualitatively similar to those in continuous wave regime, whereas depend on the ratio of pulse length to the metaslab thickness. It is shown that width and shape of transmitted fundamental and generated second harmonic pulses can be controlled by changing intensity and width of input pulses.

\begin{acknowledgements}
This material is based upon work supported in part by the U. S. Army Research Laboratory and the U. S. Army Research Office under grant number W911NF-14-1-0619, by the National Science Foundation under grant number ECCS-1346547 and by the Russian Foundation for Basic Research under grant RFBR 15-02-03959A. We thank I. S. Nefedov, A. E. Boltasseva  and V. M. Shalaev for inspiring inputs.

\end{acknowledgements}



\end{document}